\newcommand{\ro}{$R_{\rm{Freq}}$}
\newcommand{\rf}{$R_{\rm{Fourier}}$}
\newcommand{\rx}{$R_{x_1}$}

\newcommand{\project}[1]{\textsl{#1}}
\newcommand{\gaia}{\project{Gaia}}

\documentclass[a4paper,fuseAMS,leqn,usenatbib]{mnras}

\usepackage{graphicx}
\usepackage{amssymb}
\usepackage{amsmath}
\usepackage{times}
\usepackage{subfigure}
\usepackage[T1]{fontenc}
\usepackage{ae,aecompl}
\usepackage{textcomp}

\usepackage{hyperref}

\title[Constraining the Milky Way bar]{ Dynamically constraining the length of the Milky Way bar }

 \author[Lucey, et al. 2022]{Madeline~Lucey$^{1}$\thanks{E-mail:m\_lucey@utexas.edu}, 
 Sarah Pearson$^{2}$\thanks{Hubble Fellow}, 
 Jason A. S. Hunt$^{3}$,
 Keith Hawkins$^{1}$, 
 \newauthor
 Melissa Ness$^{3,4}$, 
 Michael S. Petersen$^5$,
 Adrian M. Price-Whelan$^{3}$, and
 \newauthor
Martin D. Weinberg$^6$
\\
$^{1}$Department of Astronomy, The University of Texas at Austin, 2515 Speedway Boulevard, Austin, TX 78712, USA \\
$^{2}$Center for Cosmology and Particle Physics, Department of Physics, New York University, 726 Broadway, New York, NY 10003, USA \\ 
$^{3}$Center for Computational Astrophysics, Flatiron Institute, 162 5th Avenue, New York, NY 10010, USA \\
$^{4}$Department of Astronomy, Columbia University, 550 West 120th Street, New York, NY, USA \\
$^{5}$CNRS and Sorbonne Universite, UMR 7095, Institut d’Astrophysique de Paris, 98 bis Boulevard Arago, F-75014 Paris, France \\
$^{6}$Department of Astronomy, University of Massachusetts at Amherst, 710 N. Pleasant St., Amherst, MA 01003\\
}

\date{Accepted . Received ; in original form }

\pubyear{2022}

\begin{document}
\label{firstpage}
\pagerange{\pageref{firstpage}--\pageref{lastpage}}
\maketitle

\begin{abstract}

 We present a novel method for constraining the length of the Galactic bar using 6D phase space information to directly integrate orbits. We define a pseudo-length for the Galactic bar, named \ro, based on the maximal extent of trapped bar orbits. We find the \ro\ measured from orbits is consistent with the \ro\ of the assumed potential only when the length of the bar and pattern speed of said potential is similar to the model from which the initial phase-space coordinates of the orbits are derived. Therefore, one can measure the model's or the Milky Way's bar length from 6D phase-space coordinates by determining which assumed potential leads to a self-consistent measured \ro. When we apply this method to $\approx$210,000 stars in APOGEE DR17 and \gaia\ eDR3 data, we find a consistent result only for potential models with a dynamical bar length of $\approx$3.5 kpc. We find the Milky Way's trapped bar orbits extend out to only $\approx$3.5 kpc, but there is also an overdensity of stars at the end of the bar out to 4.8 kpc which could be related to an attached spiral arm. We also find that the measured orbital structure of the bar is strongly dependent on the properties of the assumed potential.

\end{abstract}


\section{Introduction}

\label{sec:Introduction}

Stellar bars are non-axisymmetric, elongated structures in the inner parts of disk galaxies. More than 30\% of massive disk galaxies ($\rm{M_*}$ $>$ $10^{10}$ $\rm{M_{\odot}}$) in the local Universe host strong stellar bars \citep{Sellwood1993,Masters2011,Gavazzi2015}. Although many open questions remain concerning how stellar bars form and evolve, it is clear that they play pivotal roles in the secular evolution of disk galaxies \citep{Debattista2004,Athanassoula2005}.

The Milky Way hosts a stellar bar at its center, which was originally discovered from near-infrared emission \citep{Blitz1991,Weiland1994} and gas kinematics  \citep{Binney1991,Peters1975}. Stellar kinematics demonstrate that most of the mass in the inner Galaxy participates in the bar structure \citep{Howard2009,Shen2010,Ness2013b,Debattista2017}. Currently, it is under debate whether a small pressure-supported component distinct from the disk or halo (e.g., a classical bulge) overlaps with the bar \citep{Kunder2020,Arentsen2020,Lucey2021}. It has also been discovered that the center of the MW has an X-shaped structure \citep{Nataf2010,McWilliam2010,Ness2012,Wegg2013,Ness2016b}, which is characteristic of a boxy/peanut-shaped (B/P) bulge and consistent with simulations and observations of barred galaxies \citep{Combes1990,Athanassoula2005,Martinez-Valpuesta2006,Bureau2006,Laurikainen2014}.

The discovery of the MW's stellar bar provides the unique opportunity to study a bar in exquisite detail using resolved stars. However, the high levels of variable extinction have historically made the MW's bar difficult to observe \citep{Nataf2013}. Parameterizing the MW bar's mass, length and pattern speed is essential for many studies of MW dynamics. The bar greatly influences the perturbative kinematics of the Galactic disk, including in the Solar neighborhood \citep{Dehnen2000,Minchev2010,Antoja2018,Hunt2018,Fujii2019}. The Galactic bar can also impact the structure of stellar streams in the halo and the ability to interpret dark matter substructure signatures in the streams \citep{Price-Whelan2016b,Hattori2016,Pearson2017,Erkal2017,Banik2019,Bonaca2020}.

There have been a number of efforts to map the three dimensional structure of the bulge/bar region of the Galaxy, primarily using star counts \citep{Saito2011,Wegg2013}. However, the existence of a long ($\approx$4 kpc) bar, discovered initially by \citet{Hammersley1994}, has led to controversy on whether it is a separate structure from the B/P bulge \citep{Hammersley2000,Lopez-Corredoira2007,Cabrera-Lavers2007,Cabrera-Lavers2008,Martinez-Valpuesta2011}. Using the 3D number density of red clump giants from VVV, UKIDSS, GLIMPSE, and 2MASS data, \citet{Wegg2015} demonstrated that the long bar is the extension of the B/P bulge and they are in fact one unified structure. Furthermore, \citet{Wegg2015} found that the bar has a half length of 5 kpc and is at an angle of (28-33)$^{\circ}$ from the Sun-Galactic center line. However, it is possible that spiral arms connected to the bar may cause it to appear 1-1.5 kpc longer in the number density counts than when the spiral arms are not connected \citep{Gonzalez2016,Hilmi2020}. To account for this effect, it is critical to further constrain the length of the bar using a dynamical method which can distinguish between trapped bar stars and  those whose major-axes do not participate in solid body rotation.

The pattern speed of the bar, however, is better constrained with recent estimates from multiple methods clustering around 40 km $\rm{s^{-1} kpc^{-1}}$  \citep[e.g.,][]{Wang2013,Binney2020}. Adding kinematic data from the ARGOS survey \citep{Freeman2013,Ness2013a} to the work of \citet{Wegg2015}, \citet{Portail2017} found a pattern speed of 39.0 $\pm$ 3.5 km $\rm{s^{-1} kpc^{-1}}$ using the Made-to-Measure method. \citet{Sanders2019} measured a pattern speed of 41 $\pm$ 3 km $\rm{s^{-1} kpc^{-1}}$ using a direct method derived from the continuity equation \citep{Tremaine1984}. In addition, they used proper motion data of stars within 2 kpc of the Galactic center from \gaia\ DR2 and VVV surveys. Using a similar method, \citet{Bovy2019} and \citet{Leung2022} created kinematic maps of Apache Point Observatory Galactic Evolution Experiment \citep[APOGEE; ][]{Majewski2017} and \gaia\ DR2 data \citep{Gaia2016,Gaiasummary2018} to measure a pattern speed of 41 $\pm$ 3 $\rm{s^{-1} kpc^{-1}}$ out to a distance of 5 kpc from the Galactic center. However, there are also bar pattern speed estimates as high as 60 km $\rm{s^{-1} kpc^{-1}}$ \citep[e.g.,][]{Wang2012}, leaving room for some debate.

In this paper, we take advantage of the 6D phase-space measurements at the center of our Galaxy to directly integrate the orbits of stars located in the Milky Way's bar. Specifically, we develop a novel method for constraining the MW's bar length and pattern speed. To verify our method, we use N-body simulations and compare the maximal extent of stars in the bar measured from the orbits to that of the potential model used to calculate the orbits. We find that these lengths are only consistent when the initial positions and velocities of the star particles come from a distribution similar to the potential in which the orbits are integrated. We test $\approx$60 different MW bar potential models, by integrating APOGEE/\gaia\ data within these potentials and determining whether the retrieved maximal extent is consistent with the given potential model. In Section \ref{sec:models}, we describe the simulations we use to validate our method while in Section \ref{sec:data} we describe the observations used to constrain the MW's bar. We describe the method and verify its precision and accuracy in Section \ref{sec:method}. Next, we apply our method to the MW data in Section \ref{sec:MW} and discuss the different methods for measuring bar lengths in Section \ref{sec:discuss}. Last, we present our conclusions in Section \ref{sec:conclu}.

\section{Simulations}
\label{sec:models}

\begin{figure*}

    \includegraphics[width=\linewidth]{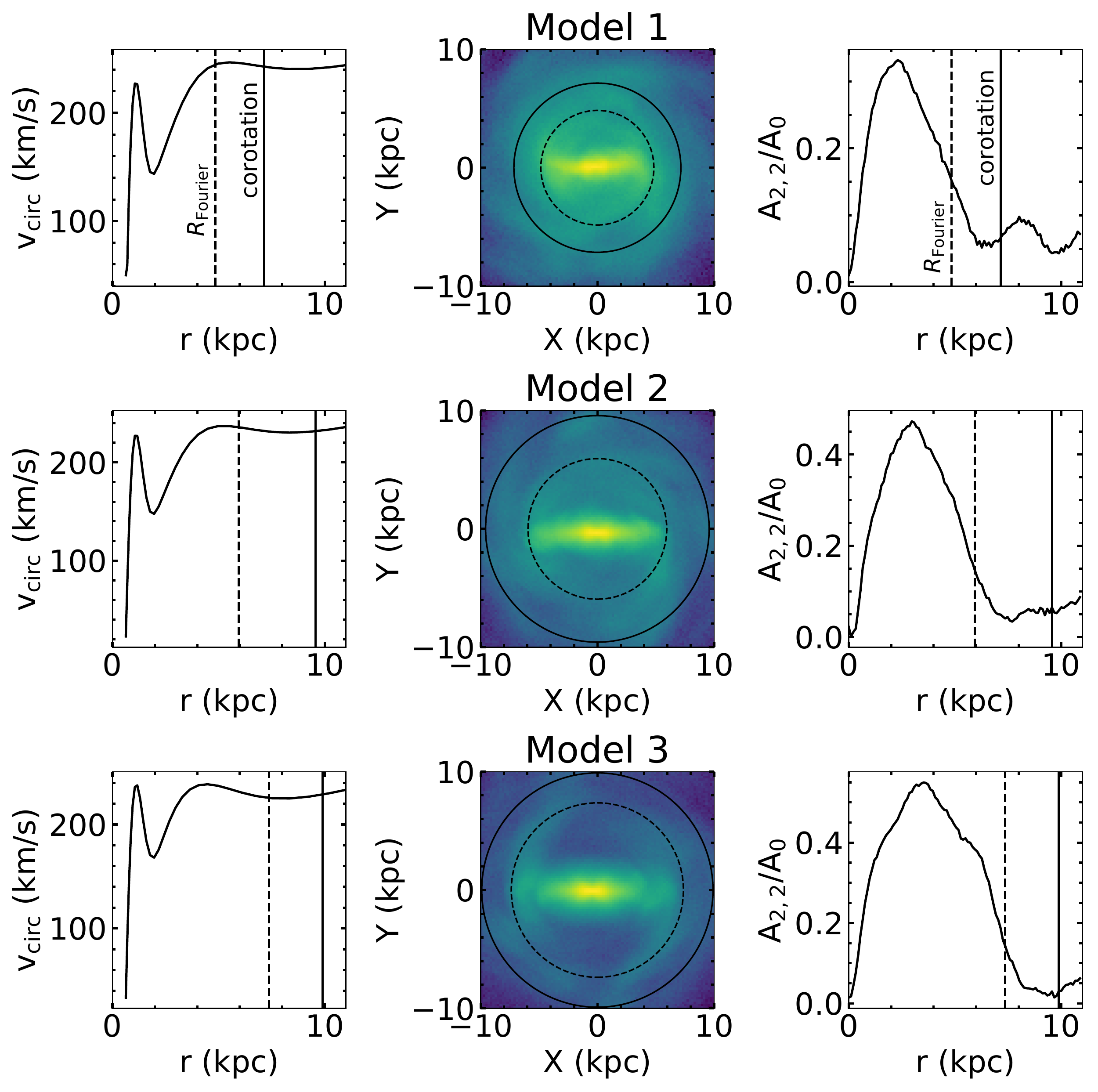}
    \caption{Properties of the three primary N-body snapshots that we utilize to validate our method. The leftmost column shows the circular velocity curves for each of the three models. In the rightmost column, we show the $A_{2,2}/A_0$ ratio (Equation \ref{eq:a2}) as a function of galactic radius. We define the bar length (\rf) as the radius at which $A_{2,2}/A_0$=0.15. In each plot, the bar length is shown as the black dashed line. We also show the corotation radius, defined as the radius where $\Omega_{\rm{b}}=v_{circ}(r)/r$, as a black  solid line. In the center column, we show face-on images of each model with circles marking \rf\ (black dashed) and corotaion radii (black solid). \rf\ visually matches the drop in number density.  }
    \label{fig:a2}
\end{figure*}

We make use of two Milky Way-like N-body simulations from the literature. The primary simulation we use (hereafter Galaxy A) is a reproduction of the MWP14-3 model from \citet{Bennett2021b}. We also make use of another Milky Way-like simulation (hereafter Galaxy B) from \citet{Tepper-Garcia2021} to further validate our method and compare to Milky Way observational data. 
 Specifically, we extract potentials from the mass distributions of 29 unique snapshots of each simulation. From Galaxy A we also extract three sets of 10,000 random initial positions and velocities of disk star particles from three different snapshots to use as initial phase-space coordinates for orbits. We call these snapshots Model 1, Model 2, and Model 3. Since we know the true underlying mass distribution for these initial phase-space coordinates, we can use them to test our method.

 Face-on images of the three snapshots used to extract initial positions and velocities are shown in the middle column of Figure \ref{fig:a2}. Model 1, 2, and 3 correspond to the snapshots that are 2.94, 3.92, and 4.90 Gyr into the simulation's evolution, respectively. Over this time, the galactic bar grows and slows. Similar to previous work \citep[e.g.,][]{Athanassoula2002,Zana2018,Rosas-Guevara2020,Rosas-Guevara2021}, we use the $m=2$ mode of the Fourier decomposition of the face-on stellar surface density to estimate the length of the bar in number density in order to compare to our dynamical estimate. We determine the Fourier components:
\begin{equation}
    A_m(r)=\frac{1}{\pi}  \int_{0}^{2\pi} \Sigma(r,\theta)cos(m\theta)\,d\theta, \hskip0.05\columnwidth  m=0,1,2,...
\end{equation}
and 
\begin{equation}
    B_m(r)=\frac{1}{\pi}  \int_{0}^{2\pi} \Sigma(r,\theta)sin(m\theta)\,d\theta, \hskip0.05\columnwidth  m=1,2,...
\end{equation}
where $\Sigma(r,\theta)$ is the stellar surface density. We then define
\begin{equation}
A_{2,2}/A_0=\sqrt{A_2^2+B_2^2}/A_0 
\label{eq:a2}
\end{equation} 
and calculate it as a function of $r$ using equal width annuli of $\Delta r$= 0.10 kpc. Similar to \citet{Rosas-Guevara2021}, we use the radius at which $A_{2,2}/A_0$=0.15 as our estimate for the bar length, which is shown as the black dashed in all panels of Figure \ref{fig:a2}.  In this work, we denote bar lengths measurements derived using this method as \rf.  The choice of value to use for $A_{2,2}/A_0$ can vary \citep[e.g.,][]{Athanassoula2002}. Arbitrarily, we could also use the radius at which $A_{2,2}/A_0=0.20$ which would shorten the bar length estimate.  For further discussion on the various methods for measuring bar lengths see section \ref{sec:discuss}.

We calculate the bar's pattern speed, $\Omega_{\rm{b}}$, by measuring the change in the $m=2$ phase angle ($\phi_2 = \frac{1}{2}\rm{tan}^{-1}(B_2/A_2)$) between simulation outputs which are 9.79 Myr apart. Here we calculate $\rm{B_2}$ and $\rm{A_2}$ between radii of 1 to 5 kpc. In the leftmost column of Figure \ref{fig:a2}, we plot the circular velocity curve for each of the snapshots. We also plot the corotation radius, where $\Omega_{\rm{b}}=v_{\rm{circ}}(r)/r$, as a black solid vertical line. Consistent with expectations, \rf\ is shorter than the corotation radius \citep{Contopoulos1980}. Specifically, we find that \rf\ is 74\%, 68\%, and 82\% of the corotation radius for Models 1, 2, and 3, respectively, making the bar in Model 1 and 3  dynamically fast, while Model 2's bar is dynamically slow \citep{Debattista2000}. 

In addition to these three models, we also use another 26 (29 in total) snapshots from Galaxy A, to create a variety of gravitational potentials with different bar lengths. These potentials are each 97.9 Myr apart, starting after bar formation, while Models 1, 2, and 3 are 979 Myr apart. We name each snapshot from which we extract a potential based on the number of years (in units of 97.9 Myr) the simulation has evolved past Models 1, 2, or 3. For example, Model 1.2 corresponds to 195.8 Myr ($2\times97.9$ Myr) after Model 1. We use these as test potentials to determine how bar orbits are impacted by potentials with different bar strengths and lengths. Furthermore, we also integrate the APOGEE/\textit{Gaia} data in these potentials in order to determine which potential best fits the data.

The initial conditions for the Galaxy A simulation are derived from the \texttt{GALPY} potential \texttt{MWPotential2014} \citep{Bovy2015} and set up with the \texttt{GALIC} package \citep{Yurin2014}. However, \citet{Bennett2021a} found that the halo mass of \texttt{GALPY}'s  \texttt{MWPotential2014} needs to be increased in order to produce realistic asymmetries (e.g., spiral arms and a stellar bar) in simulations. Therefore, Galaxy A (MWP14-3) has a virial halo mass of $M_h=1.4 \times 10^{12} M_{\odot}$ which is twice as heavy as the halo in \texttt{GALPY}'s \texttt{MWPotential2014}. It also has a disk scale height of 0.28 kpc and disk scale length of 3.0 kpc. In total there are $\approx$9.3 million particles with 3,337,406 particles in dark matter halo, 5,000,000 particles in the disk and 996,403 particles in the bulge. The dark matter particles have masses of $3.48\times 10^5 M_{\odot}$. While the disk and bulge star particles have masses of $4.50 \times 10^3 M_{\odot}$ and $1.34 \times 10^4 M_{\odot}$, respectively. In total, this simulation is evolved for $\approx$4.99 Gyr.  

Although it is also set up to mirror the Milky Way, Galaxy B has significantly different initial conditions than Galaxy A. For a complete description of Galaxy B, we refer the reader to \citet{Tepper-Garcia2021}. In short, Galaxy B is set up using the Action-based GAlaxy Modelling Architecture software package \citep[\texttt{AGAMA;}][]{Vasiliev2019}. Compared to Galaxy A, Galaxy B has a lighter virial halo mass of $M_h=1.18\times 10^{12}M_{\odot}$. Furthermore, the disk of Galaxy B has a shorter scale height (0.3 kpc) and length (2.6 kpc) compared to Galaxy A as well as a large velocity dispersion ratio ($\sigma_R/\sigma_z=2$).

Galaxy A and B are both simulated for about 5 Gyr and each have their bars fully formed between 2 and 2.5 Gyr. However, the bar in Galaxy A significantly grows and slows over the simulation whereas the bar in Galaxy B stays rather stable (see x-axis of Figure \ref{fig:len}). Using an iso-density contour, \citet{Tepper-Garcia2021} measure a bar length of $\approx$4.5 kpc, and pattern speed of $\approx$40 km/s/kpc. 

As for Galaxy A, we use a total of 29 snapshots from Galaxy B to create gravitational potentials. Similar to Galaxy A, these potentials are 95 Myr apart, starting after bar formation. However, as the bar is more stable in this simulation, these potentials have a much smaller range in bar lengths. Importantly, these potentials allow us to ensure that our method works across different simulations and is not dependent on  properties of the simulated galaxies' potentials other than the bar length and pattern speed. Given that the properties of bars are thought to be heavily impacted by dark matter halos \citep[e.g.,][]{Petersen2019a,Debattista2000,Fragkoudi2021,Chiba2021,Collier2021}, it is especially important that we use two models that have different dark matter halo masses and profiles.

For each of the 58 snapshots (29 each from Galaxies A and B), we first use the mass distribution to extract the corresponding gravitational potential using the AGAMA package \citep{Vasiliev2019}. Specifically, we use a multipole expansion to represent the spherical bulge and dark matter halo components. For the disk, we use the \texttt{CylSpine} potential representation which uses azimuthal Fourier harmonics.   

\section{Data} \label{sec:data}

\subsection{APOGEE DR17 and Gaia eDR3}

\begin{figure}

    \includegraphics[width=\columnwidth]{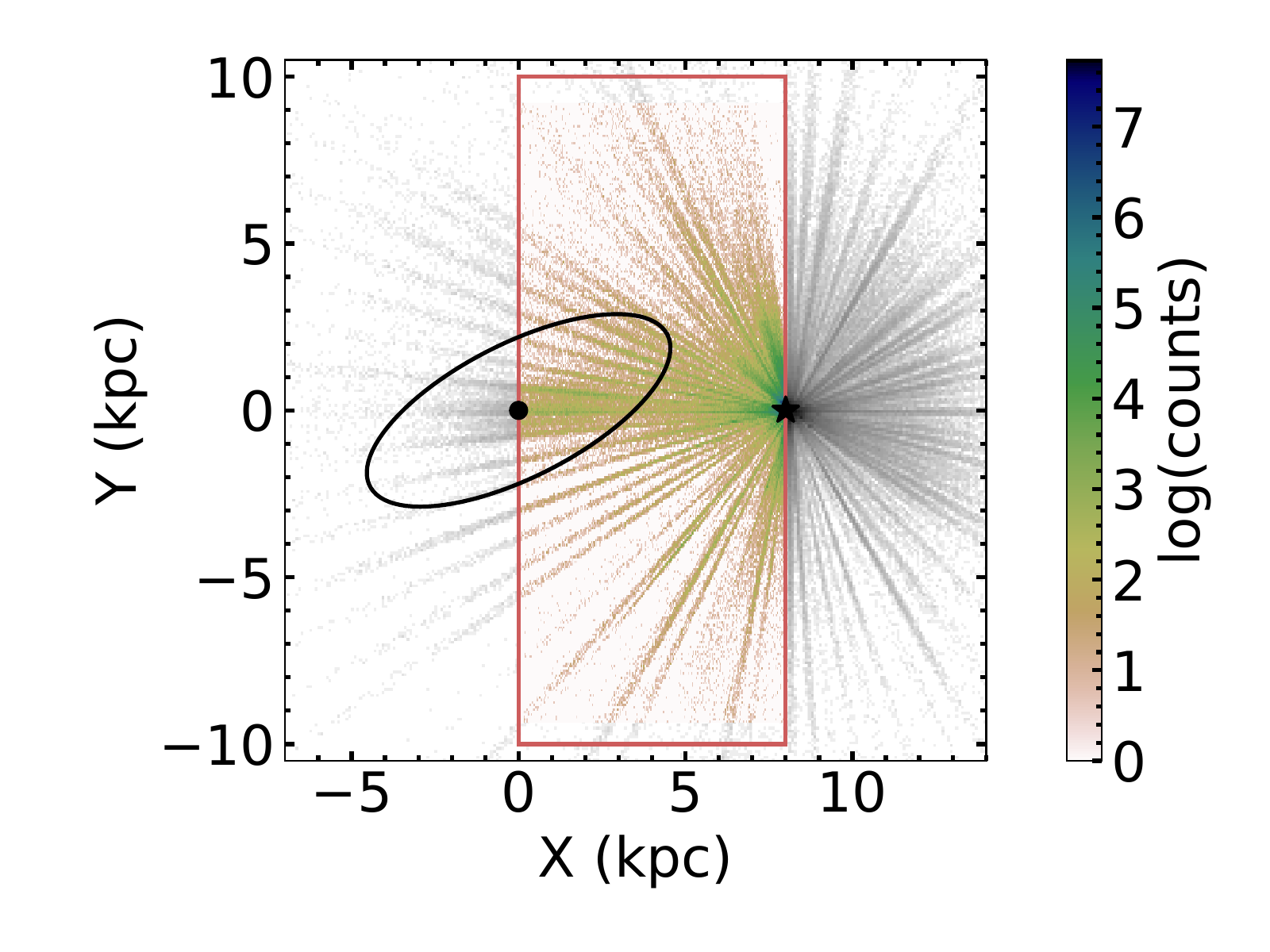}
    \caption{Galactic distribution of the astroNN catalog using APOGEE DR17 and \gaia\ eDR3 \citep{Leung2019}. The Sun is located at (8.3,0,0) kpc with the Galactic center at (0,0,0) kpc. For this work, we use stars with 0 kpc < X < 8.3 kpc and |Y|<10 kpc (shown as a red box) in order to loosely target the Galactic bar. For reference we show the  proposed bar model of \citet{Wegg2015} as an ellipse with a semi-major axis of 5 kpc and axis ratio of 0.4 rotated 27$^{\circ}$ from the Sun-Galactic center line.}
    \label{fig:data}
\end{figure}

In order to constrain the Milky Way's galactic bar, we utilize one of the largest sets of 6D positional and kinematic data available. Specifically, we use a combination of \gaia\ eDR3 data \citep{gaiaedr3} with APOGEE DR17 \citep{Abdurro'uf2021}. APOGEE is a near-infrared (1.5-1.7 $\mu$m), high-resolution (R=$\lambda/\Delta \lambda \approx$ 22,500) large spectroscopic survey \citep{Nidever2015,Wilson2019,Zasowski2017}. APOGEE DR17 contains over 657,000 stars observed using the APO 2.5m telescope \citep{Gunn2006} and the 2.5m telescope at Las Campanas Observatory (LCO) \citep{Bowen1973}. In this work, we use the line-of-sight velocities from the ASPCAP pipeline \citep{Garcia-Perez2016,Smith2021}, along with the spectrophotometric distance estimates from the astroNN catalog \citep{Leung2019} and proper motions from \gaia\ eDR3 \citep{gaiaedr3}. Combining this data gives us typical phase-space uncertainties on the order of 5\% for all components.  We also repeat our analysis with distances from the StarHorse catalog \citep{Queiroz2020} to ensure that our results are not impacted by a possible distance bias.

In Figure \ref{fig:data}, we show the Galactic distribution of the astroNN catalog that we use in our analysis. For reference, we also show the proposed bar model of \citet{Wegg2015} as an ellipse with a semi-major axis of 5 kpc and axis ratio of 0.4, rotated 27$^{\circ}$ from the Sun-Galactic center line.  As demarcated by the red box, we choose stars in between the Sun and the Galactic center, with 0 kpc< X< 8.3 kpc and |Y|<10 kpc, in order to loosely target the Galactic bar. We choose not to use stars with X < 0 kpc due to the large distance uncertainties. This selection gives us a sample of 215,869 stars for which we have 6D spectrophotometric phase-space information for use in our analysis.  We also redo our analysis only with stars that were targeted as part of APOGEE's main science program for the bulge which is based on a simple color cut criterion. This test allows us to ensure that selection function effects do not dominate our results.

\section{Using Orbit Integration to Constrain the Length of Bars} \label{sec:method}

There are a number of methods used to estimate the length of a Galactic bars. In Table \ref{tab:table1}, we outline each measurement of bar length discussed in this work. In this section, we describe in detail the method that we develop for measuring the length of bars from orbit integration of 6D phase-space observations. 

First, to separate trapped bar stars from regular disk stars, we perform fundamental frequency analysis in the rotating bar-frame where regular disk orbits will be symmetric in x and y (i.e., $\Omega_x$ = $\Omega_y$) while bar orbits may not. Next, we define a quantity based on the apocenter distribution of orbits in the bar which we call \ro, for short. Specifically, \ro\ is defined as the 99.5th percentile of the apocenter distribution of bar stars selected from frequency analysis. We perform a number of experiments to test how each input of the orbit simulations (gravitational potential, initial phase-space coordinates of particles, and pattern speed) impact the inferred \ro. We find that only when the assumed potential has a similar bar length and pattern speed to the model from which  initial 6D positions and velocities are extracted then the inferred \ro\ is equal to the assumed potential's \ro. In other words, to achieve a self-consistent \ro\ in that the inferred value matches that of the gravitational potential used to calculate it, the assumed gravitational potential must have a bar length similar to that of the initial phase-space coordinates.  We apply this method, first, to simulations in order to confirm the accuracy and precision of the results. Next, we apply the method to the \textit{Gaia} and APOGEE data set in order to constrain the length of the Milky Way's bar.

\subsection{Defining \ro}

\label{sec:define}

\begin{table*}
\caption{Descriptions of the Different Measurements of Bar Length Used in This Work }
\label{tab:table1}
\begin{tabular}{p{0.15\linewidth} | p{0.8\linewidth}}
\hline\hline
Term & Description \\
 \hline
\rf\ & Estimate of the bar length calculated from the Fourier analysis of number density counts. Specifically, the radius at which $A_{2,2}/A_0$=0.15. See Section \ref{sec:models} for further information.   \\
\\
Potential's \ro & The bar length of the potential estimated by integrating 10,000 particles from the potential model in the potential rotating at the true bar pattern speed. The bar length is then the 99.5th percentile of the apocenter distribution of bar stars selected from orbital frequency analysis.  See Section \ref{sec:define} for more information.   \\
\\
Inferred \ro & The 99.5th percentile of the apocenter distribution of bar stars selected from orbital frequency analysis. The orbits may be calculated with any combination of initial phase-space coordinates, potential model and pattern speed. See Section \ref{sec:define} for more information.    \\
\\
\rx &  Estimate of the bar length based on the maximal extent of orbits classified as $x_1$. See Section \ref{sec:discuss} for more information.   \\
\hline
\end{tabular}
\
\end{table*}

\begin{figure*}
    \centering
    \includegraphics[width=\linewidth]{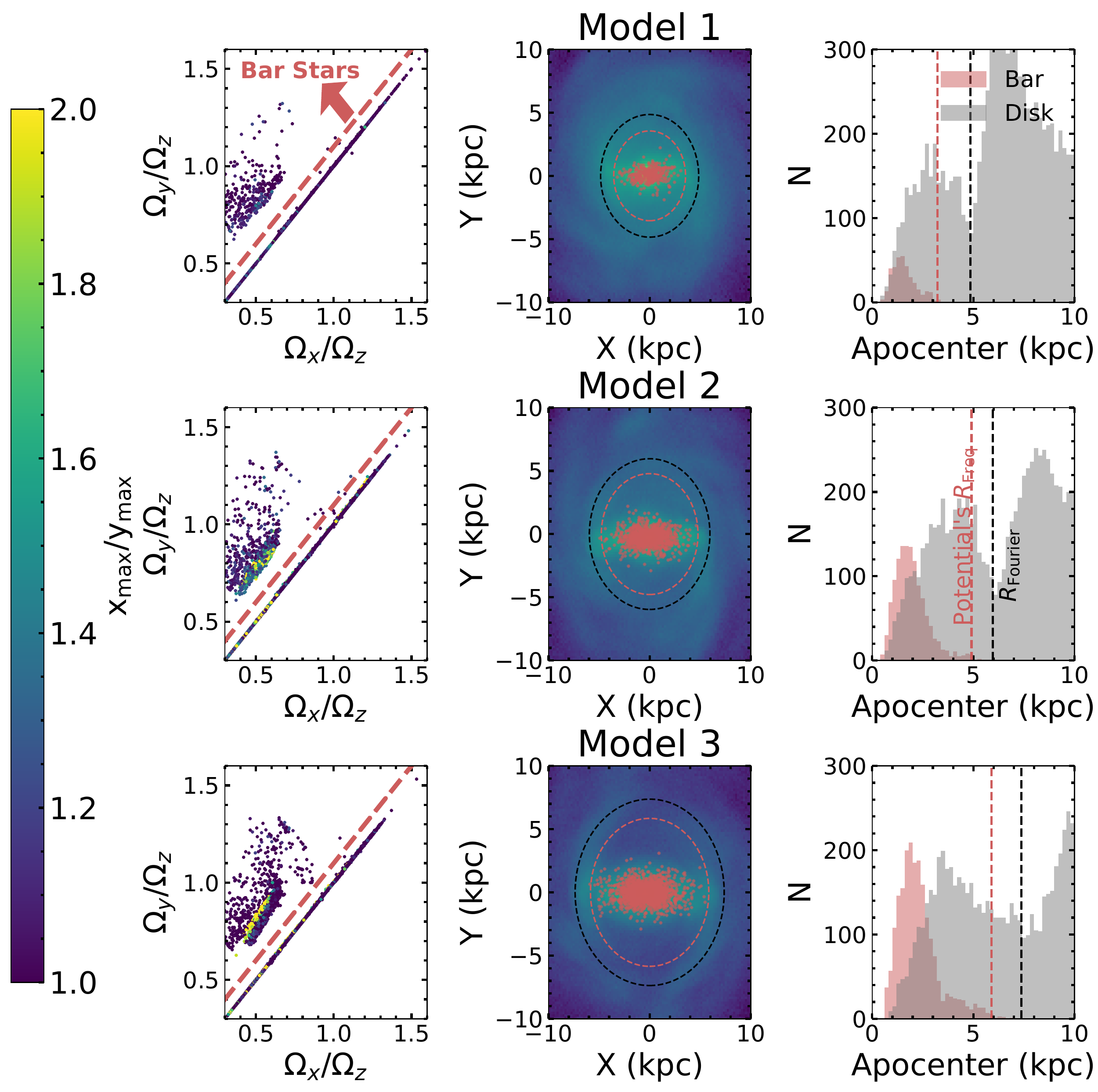}
    \caption{The selection and distribution of bar stars for three bar models. The first column shows the orbital frequency map for 10,000 disk stars randomly selected from the N-body models colored by the ratio of $\rm{x_{max}/y_{max}}$ where x is along the bar's major axis and y is along the minor axis. We show a red dashed line corresponding to $\Omega_y/\Omega_z$=$\Omega_x/\Omega_z$+0.1, as we select stars above this line as stars in the bar.  The central column shows the spatial distribution of our selected bar stars (red points) compared to the rest of the stars in the models. The last column shows the apocenter distribution from the orbits integrated in the corresponding model potential for 1 Gyr. The selected bar stars are shown in red and the rest of the 10,000 stars are shown in grey. The red dashed line correspond to the 99.5th percentile of the bar stars' apocenter distribution which we define as the potential's \ro.   }
    \label{fig:big}
\end{figure*}

In this work, we perform experiments to determine how the inferred \ro\ differs when changing the various inputs into orbit simulations. Specifically, as we are interested in applying this method to Milky Way data, we test how \ro\ changes for a given set of initial phase-space coordinates under various assumptions about the gravitational potential. For each experiment, we integrate 10,000 particles in a rotating potential for a total of 1 Gyr in timesteps of 1 Myr. We have also tried integrating the orbits for longer periods of time but do not see significant changes in our results. Although a typical bar's structure would likely evolve over 1 Gyr, assuming a stable bar for orbit integration is reasonable in our case where we are simply interested in measuring the present day bar length.

In order to calculate the inferred \ro, we first must define bar orbits. To accomplish this, we use orbital frequency analysis. Regular orbits in triaxial potentials have three fundamental frequencies (${\bf\Omega}\equiv \{\Omega_x,\Omega_y,\Omega_z\}$) which describe the periodic motion. As first demonstrated by \citet{Binney1982}, Fourier transforms can be used to recover the fundamental frequencies. Specifically, the Fourier transform gives the spectrum of orbital oscillations in the chosen coordinate system, where dominant lines tend to correspond to the fundamental frequencies, though is not always the case. Referred to as ``Numerical Analysis of Fundamental Frequencies" (NAFF), this method has been further developed and applied to galactic dynamics \citep[e.g.,][]{Laskar1993,Valluri1998,Valluri1999,Valluri2010,Price-Whelan2016a,Yavetz2021,Koppelman2021}. In this work, we use the SUPERFREQ code \citep{superfreq2,superfreq} to calculate the fundamental frequencies for each of our 10,000 stars. Specifically, we find the Cartesian fundamental frequencies ($\Omega_x, \Omega_y, \Omega_z$) in the rotating frame where the bar is stationary, which have been shown to better classify bar stars than frequencies in cylindrical coordinates \citep{Valluri2016}. Consistent with \citet{Valluri2016}, $x$ corresponds to the direction along the bar's major axis, y is along the bar's minor axis and z is the direction out of the plane. 

Disk stars on regular orbits lie along the $\Omega_x/\Omega_y= 1$ resonance line. As shown in Figure \ref{fig:big} and in \citet{Valluri2016}, bar stars tend to lie above this line with $\Omega_x/\Omega_z> 0.3$. We select stars within this region as our sample of bar stars. Specifically, we use stars above the red dashed line where $\Omega_y/\Omega_z> \Omega_x/\Omega_z+0.1$. Although this selection will certainly miss some of the bar stars, there is no contamination of the sample by disk stars. For our work, low contamination is prioritized over completeness. In Section \ref{sec:discuss}, we present a more robust method for selecting bar stars. However, this method is significantly more computationally expensive. The frequency estimates provide a fast selection of bar stars which is required given that we wish to perform the bar length calculation for a large variety of initial phase-space conditions and potentials. For this part of the analysis, we primarily wish to perform an apples-to-apples comparison between the potential's \ro\ and the inferred \ro to check for self-consistency. It is possible that the bar orbit family with the largest physical extent may vary for different bars. Therefore, missing a given family of lower-order bar orbits could potentially have disparate effects and cause an anomalously self-consistent result. However, we do not see evidence of this in our tests with simulations (see Section \ref{sec:verify}). In future work, we plan to build a faster method for classifying bar stars which will allow for a more robust analysis of the maximal extent of bar orbit families.

Now that we have isolated bar orbits, we can estimate the length of the bar based on the maximal extent of these orbits. In this work we use the 99.5th percentile instead of the maximum of the apocenter distribution in order to minimize the effect of outliers.

In each experiment, we compare the inferred \ro\ to the bar length of the potential used to calculate it. In this way, we are checking for self-consistency and testing whether the bar length of the potential is consistent with the \ro\ inferred with it. As we wish to perform an apples-to-apples comparison, we measure the bar length of the potential using the same method as \ro. To calculate the potential's \ro, we use initial phase-space coordinates of particles extracted from the given potential model. Furthermore, we rotate the potential with the correct pattern speed, corresponding to the potential's N-body snapshot, measured from Fourier analysis (see Section \ref{sec:models}). In this way, we are simulating the orbits as close as possible to the N-body snapshot on which the potential is based. Therefore, the potential's \ro\ is meant to be the ground truth \ro\ for the given potential. 

In Figure \ref{fig:big}, we demonstrate our method for defining stars in the bar and the potential's \ro. Specifically, we show results for dynamically measuring the potential's \ro\ that correspond to Models 1, 2, and 3 (see Figure \ref{fig:a2}). The leftmost column shows orbital frequency maps for Models 1, 2, and 3, respectively.  To create the frequency maps, we  plot the ratio $\Omega_x/\Omega_z$ on the x-axis and $\Omega_y/\Omega_z$ on the y-axis. We color the points by the ratio of the maximum x-position ($x_{max}$) to the maximum y-position ($y_{max}$) for each star. 

In the middle column of Figure \ref{fig:big} we show face-on images of the N-body models with the bar stars selected from the orbital frequency maps (left panels) as red points. As expected, the selected bar stars all fall within the bar region. In the rightmost column of Figure \ref{fig:big}, we plot the apocenter distribution of the orbits of these same stars (red) compared to the remaining disk stars (grey). We plot the potential's \ro\ as a red vertical dashed line. For Models 1, 2, and 3, we measure a potential \ro\ of 3.22 kpc, 4.89 kpc, and 5.88 kpc, respectively. We compare this to the \rf, which is shown as a black vertical dashed line. As shown, the potential's \ro is consistently smaller than \rf. In fact, the potential's \ro\ is between 73\% and 80\% of \rf\ for Models 1, 2, and 3. However, it is important to note that \rf\ is known to overestimate the length of bars especially in the case of attached spiral arms \citep{Hilmi2020,Petersen2019b}. We refer the reader to Section \ref{sec:discuss} for further comparisons and discussion of methods for estimating the bar length.

For the potential's \ro, we are always using particles extracted from the N-body model whose mass distribution is the basis of the potential. This is not necessarily true for the inferred \ro.  In the next section, we test how the inferred \ro\ changes from the potential's \ro\ as we change the potential, but fix the initial phase-space coordinates. It is reasonable to expect that inferred \ro\ is approximately the same as the potential's \ro\ when the potential is similar to the Model from which the initial positions and velocities are extracted. However, the power of our method comes from the fact that the inferred \ro\ and potential's \ro\ are approximately the same \textit{only} when the potential is similar to the Model from which the initial positions and velocities are extracted as we show in Section \ref{sec:verify}.

\subsection{Verifying the Method with Simulations}
\label{sec:verify}

 \begin{figure*}
    \centering
    \includegraphics[width=\linewidth]{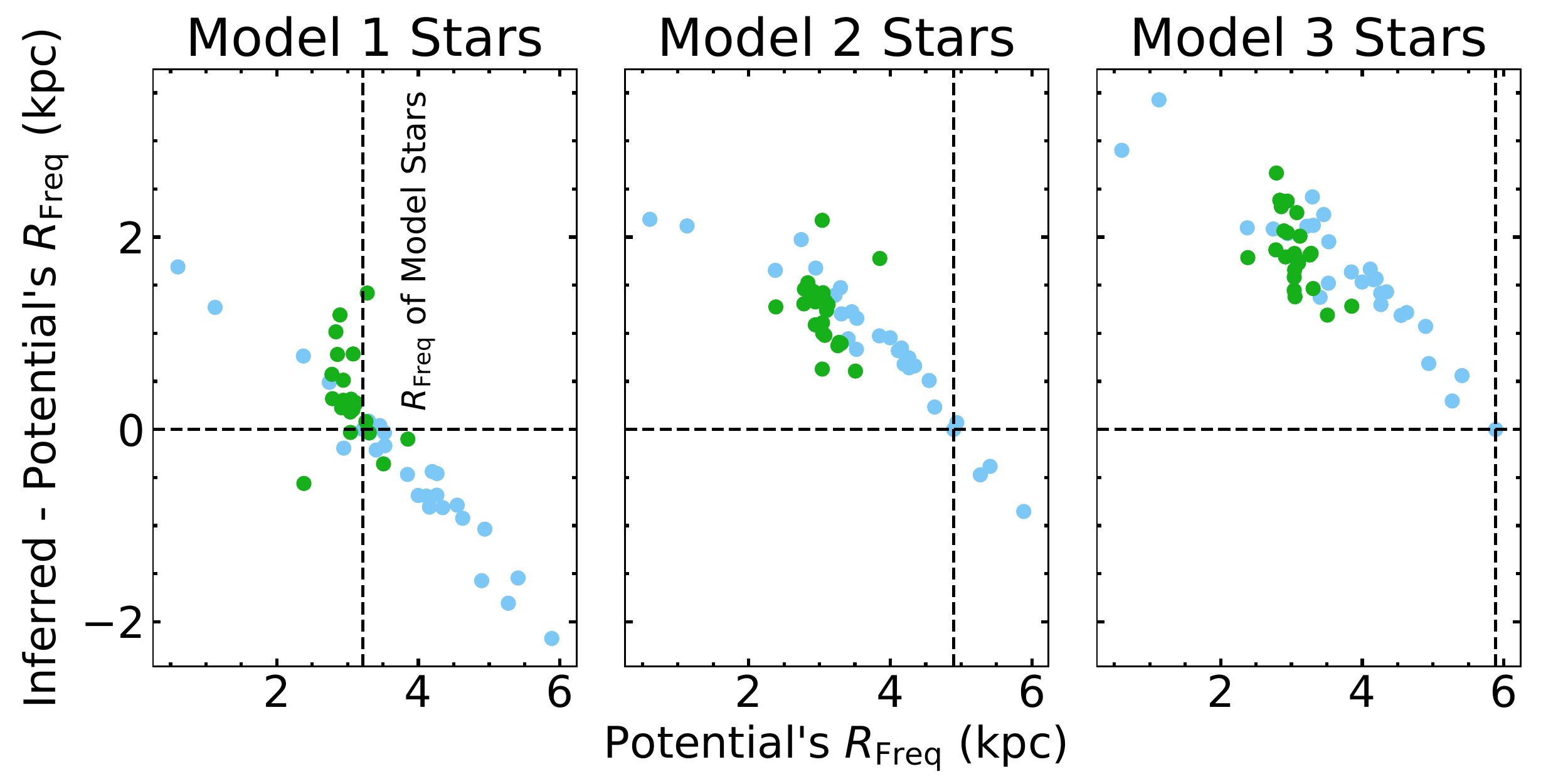}
    \caption{ The change in the inferred \ro\ compared to the \ro\ of the potential. The inferred \ro\ is calculated from integrating 10,000 stars extracted from Model 1 (left), Model 2 (center) and Model 3 (right) for 1 Gyr in potentials with different bar lengths. Results using potentials based on the 29 Galaxy A snapshots are shown in light blue, while results using potentials based on the 29 Galaxy B snapshots are in green. The pattern speed is fixed to that of the model's bar from which the initial phase-space coordinates are extracted (Model 1, 2, or 3). The vertical black dashed line shows the \ro\ of Model 1, 2 or 3, respectively. }Note, we get a consistent result (y-axis value $\approx$ 0) only when the stars' 6D position and velocity data is consistent with the potential used.  
    \label{fig:len}
\end{figure*}

\begin{figure*}
    \centering
    \includegraphics[width=\linewidth]{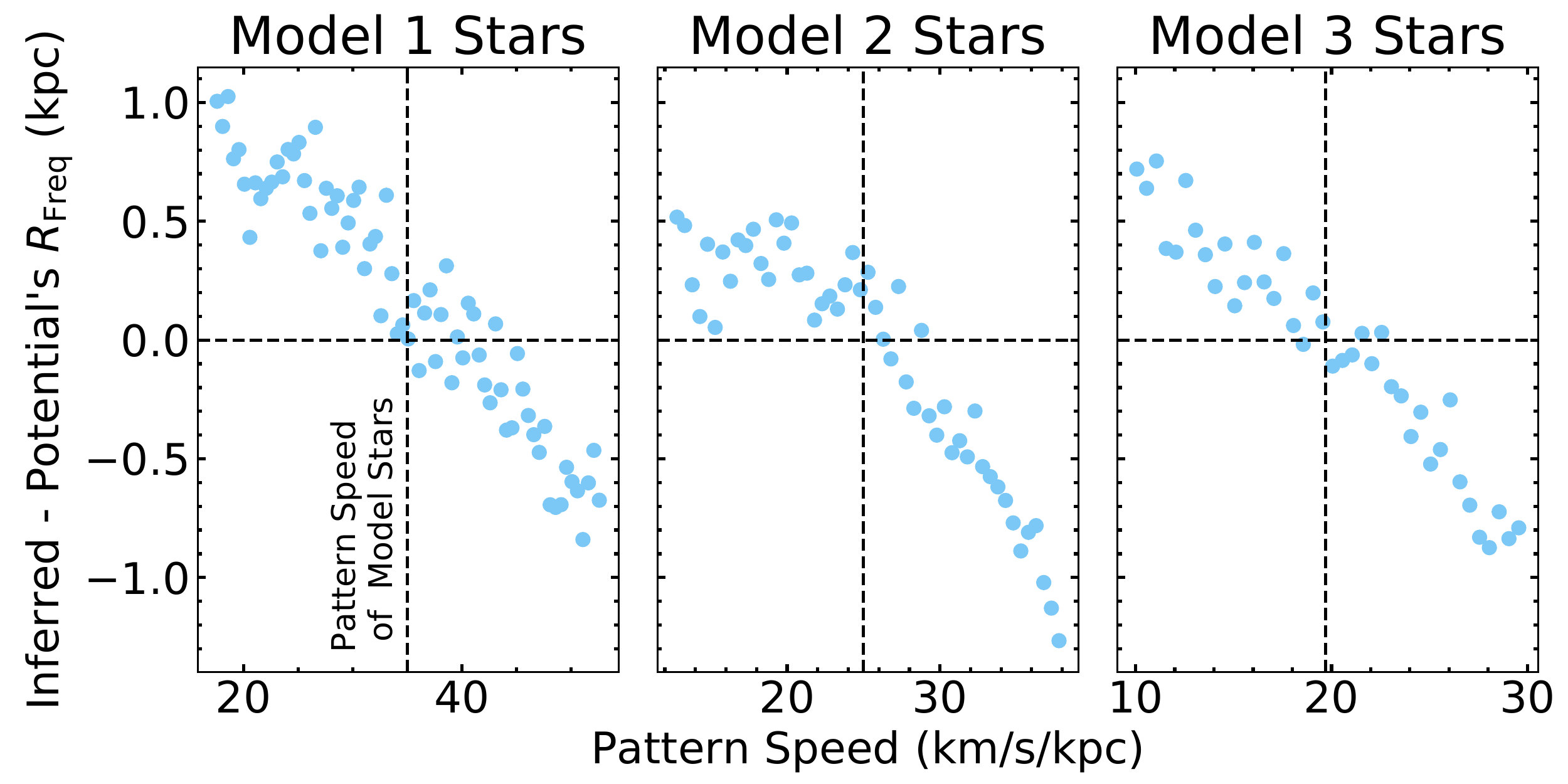}
    \caption{The effect of changing the potential's  pattern speed on the inferred \ro. In this plot, we use the same stars as in Figure \ref{fig:len}, but we only use the corresponding Model's potential. We integrate 10,000 stars in the same potential for 1 Gyr using different pattern speeds each time and compare the inferred \ro\ from the orbits to the potential's \ro. The vertical dashed line corresponds to the pattern speed calculated from the Model stars. In general, increasing the pattern speed at which the potential is rotated shortens the inferred \ro. The inferred \ro\ is consistent with the potential's \ro, only when the pattern speed is consistent with the Model stars.}
    \label{fig:ps}
\end{figure*}

\begin{figure*}
    \centering
    \includegraphics[width=\linewidth]{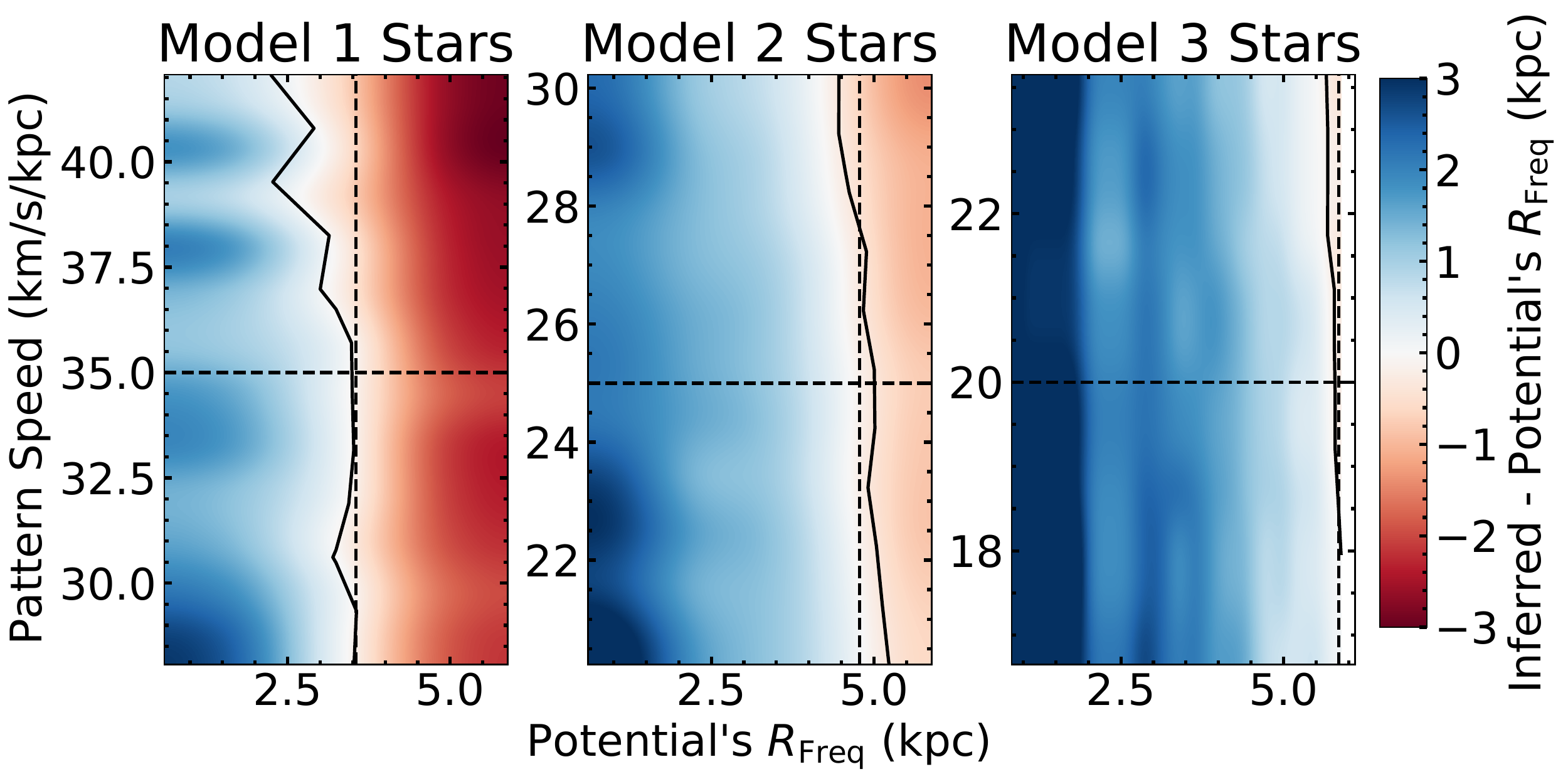}
    \caption{The difference between the inferred \ro\ and the potential's \ro\ as a function of the pattern speed and potential's \ro. This figure is similar to Figure \ref{fig:len}, except we also change the pattern speed (y-axis) by up to $\pm$ 20\%. The color corresponds to the difference between the inferred \ro\ and the potential's \ro, where red means the potential's \ro\ $>$ the inferred \ro\, white is where the potential's \ro $\approx$ the inferred \ro\ and blue means the potential's \ro\ $<$ the inferred \ro. The black line indicates where the potential's \ro= the inferred \ro. The black and grey vertical dashed lines  mark the \ro\ and associated uncertainty of the model from which the stars' initial positions and velocities were extracted, while the horizontal black dashed line indicates its pattern speed.  Even when the pattern speed is different than that of the stars by $\lesssim$ 20\%, we still generally find a consistent inferred \ro\ only when the potential is consistent with the star's initial positions and velocities. }
    \label{fig:wrong}
\end{figure*}


In this section, we perform three types of experiments. First, we test how the inferred \ro\ changes compared to the potential's \ro\ as we use potentials with varying bar length, but we fix the initial phase-space coordinates. We also hold the pattern speed of the potential constant at the value corresponding to the bar's pattern speed in the Model from which the initial phase-space coordinates are extracted. We show the results of these experiments for Model 1, 2 and 3 stars in Figure \ref{fig:len}. In our next type of experiment, we hold the potential model and the initial phase-space coordinates constant but change the potential's pattern speed. The results of these experiments are shown in Figure \ref{fig:ps}. In our third type or experiment, which is shown in Figure \ref{fig:wrong}, we vary both the pattern speed and the potential model, holding only the initial phase-space coordinates constant. 

In Figure \ref{fig:len}, we show the change in the inferred \ro\ for Model 1, 2, and 3 stars as they are integrated in potentials with different \ro . On the y-axis, we show the inferred \ro\ compared to the \ro\ of the potential.  We also show the potential's \ro\ on the x-axis. The left, center and right panels shows results for stars with initial positions and velocities extracted from Model 1, 2 and 3, respectively. We show the potential's \ro\ of Model 1, 2, and 3 as black vertical dashed lines with Models 1, 2 and 3 having a potential \ro\ of $\approx$3.22 kpc, 4.89 kpc, and 5.88 kpc, respectively (see Figure \ref{fig:big}). It should be noted that here we assume the pattern speed is known and we rotate the potential with the pattern speed calculated from the corresponding Model stars.

We find that only when the potential's bar length is similar to the bar length of the model from which the initial positions and velocities are extracted do we measure a consistent inferred \ro, i.e., the difference between the inferred and potential's \ro\ is $\approx$0 kpc. That is to say, if the inferred \ro\ is significantly different from the \ro\ of the potential used to calculate the orbits, then we know that said potential is not representative of the initial 6D positions and velocities of the stars since it does not give a consistent result. Therefore, we can apply this to the Milky Way and constrain the Galactic bar's potential by testing which potential model provides an inferred \ro\ that is the same as the potential's \ro\ using APOGEE and \gaia\ data as our initial positions and velocities. However, first it is important to investigate the impact of different pattern speeds on these results given that the Milky Way's pattern speed is uncertain, although somewhat well-constrained.

In addition to the potential's bar length, we find that varying the potential's bar pattern speed also impacts the inferred \ro. Similar to Figure \ref{fig:len}, we recalculate the inferred \ro\ for Model 1, 2, and 3 stars, but instead of using potentials from different snapshots we assume the bar length is known. Specifically, we use the corresponding Model's potential and simply vary the pattern speed at which the potential is rotated as the orbits are integrated. We show the impact of varying the pattern speed on the inferred \ro\ in Figure \ref{fig:ps}. Similar to Figure \ref{fig:len}, we have the difference between the inferred and potential's \ro\ on the y-axis. However, now we have the pattern speed on the x-axis. The vertical dashed line corresponds to the pattern speed of the bar in the respective Model 1, 2, and 3 snapshots (see Section \ref{sec:models} for calculation).  In general, an increase in the pattern speed leads to a shorter inferred \ro, while a decrease in the patter speed lengthens the inferred \ro.  However, we need to determine how this behavior changes when the potential does not necessarily represent the model of the initial phase-space coordinates. 


At this point, we have tested how the inferred \ro\ changes for a given set of initial phase-space coordinates when we vary either the potential's bar length or pattern speed. However, we have only varied one of these parameters at a time while we fixed the other to match that of the initial phase-space coordinate's model. Here, we test whether agreement between the potential and pattern speed will always lead to an inferred \ro\ consistent with the potential's, independent of the initial phase-space coordinates. Specifically, we perform the same experiment as in Figure \ref{fig:len}, but instead of fixing the pattern speed to the value corresponding to the initial phase-space coordinate's models, we vary it so that it matches the potential model's bar pattern speed.  We find that the initial phase-space coordinates play a major role in the inferred \ro\ and agreement between the potential and pattern speed does not always lead to an inferred \ro\ that matches the potential. Consistent with Figure \ref{fig:ps}, a faster or slower pattern speed can sometimes shorten or lengthen the inferred \ro\ to match the potential. Therefore, for each potential we could likely find a pattern speed that would shorten or length the inferred \ro\ to match that of the potential. Thus, it is crucial to have constraints on the pattern speed in order to determine the most consistent potential model.


In Figure \ref{fig:wrong}, we test whether we can determine the correct potential model if the pattern speed is known to within 20\%. Specifically, we recalculate the inferred \ro\ for Model 1, 2, and 3 stars using potentials with different length bars \textit{and} different bar pattern speeds.  The pattern speed is shown on the y-axis, while the potential's \ro\ is on the x-axis. The color corresponds to the difference between the inferred and potential's \ro\ where blue means the inferred \ro\ $>$ the potential's \ro, white is where the inferred \ro\ $\approx$ the potential's \ro and red means the potential's \ro $>$ the inferred \ro. The black line indicates where the inferred \ro\ = the potential's \ro. The vertical black lines correspond to the \ro\ of the initial phase-space coordinates' Models. The horizontal black dashed lines correspond to the Model's bar pattern speed. As the difference in estimates for the Milky Way's pattern speed are $\approx$20\% \citep{Bovy2019}, we wish to investigate the behavior of the inferred \ro\ when the assumed pattern speed is incorrect by up to 20\%. We note that even the fastest pattern speed for a given Model does not decrease the corotation radius below the Model's $\rm{A_2/A_0}$ bar length which would cause the bar to become ultrafast and violate our theoretical understanding of bars \citep{Contopoulos1980,Contopoulos1981,Buta2009,Vasiliev2015}. 

Even when the pattern speed does not match the initial positions and velocities of the stars, in general, we still find that we retrieve a consistent inferred \ro\ only when the assumed potential's bar length is similar to the bar length of the N-body snapshot from which the initial positions and velocities of the stars were taken.  Consistent with Figure \ref{fig:ps}, we do see that at the highest pattern speeds, the inferred \ro\ begins to decrease. However, this effect is generally small compared to the uncertainties on \ro\ when the pattern speed is within 20\% of the Model's pattern speed. For the shorter bar in Model 1 (left panel),  the faster pattern speed does lead to a  potential model with a shorter bar being most consistent with the inferred \ro. Therefore, we test a slow and fast bar model for the APOGEE data in Section \ref{sec:MW}.  In general, it is possible to determine the bar length that corresponds to the initial positions and velocities of the stars if the bar's pattern speed is known to within 20\%. As the Milky Way's bar pattern speed is known to within 20\% \citep{Bovy2019}, we conclude that we can determine which bar potential is most consistent with the APOGEE and \gaia\ data. 

\section{Constraining the Milky Way's Bar Length}

\label{sec:MW}

\begin{figure*}
    \centering
    \includegraphics[width=\linewidth]{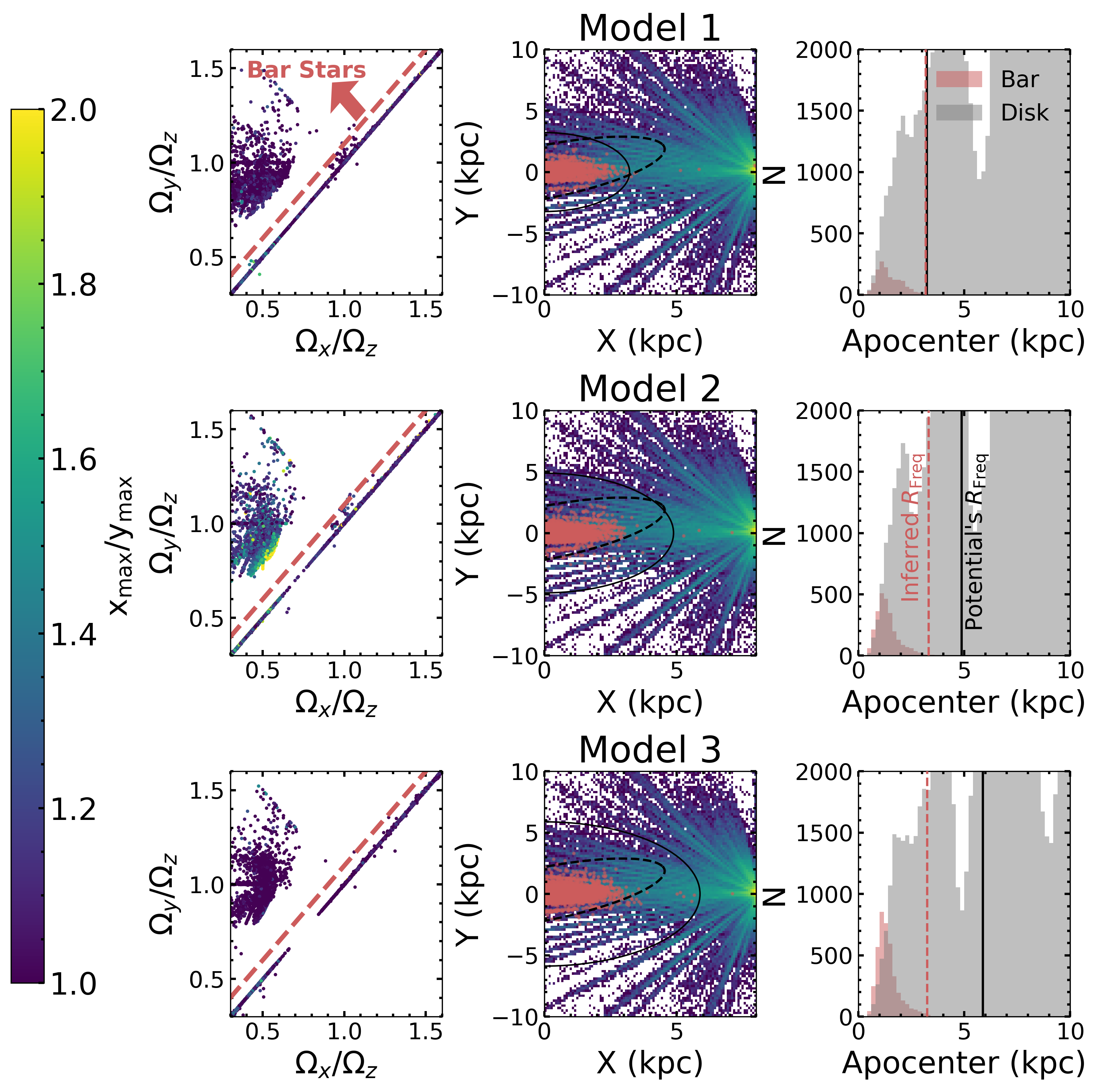}
    \caption{The selection and distribution of bar stars for the APOGEE/\gaia\ data integrated in three different potentials with different bar lengths. We assume a bar pattern speed of 41 km/s/kpc for the orbit calculation. The first column shows the orbital frequency map for $\approx$210,000 stars selected according to Figure \ref{fig:data} colored by the ratio of $\rm{x_{max}/y_{max}}$ where x is along the bar's major axis and y is along the minor axis. We also show a red dashed line corresponding to $\Omega_y/\Omega_z = \Omega_x/\Omega_z+0.1$, as we select stars above this line as bar stars.  The central column shows the spatial distribution of our selected bar stars (red points) compared to the rest of the stars. We also show the same bar model as in Figure \ref{fig:data} (dashed black line) as well as a circle with radius equivalent to the potential's \ro\ 
    (solid black line). The last column shows the apocenter distribution from the orbits integrated in the corresponding model potential for 1 Gyr. The selected bar stars are shown in red and the rest of the stars are shown in grey. We also show the \ro\ of the potential as a black solid line. The red dashed line corresponds to the 99.5th percentile of the bar stars' apocenter distribution i.e., the inferred \ro. }
    \label{fig:apoig}
\end{figure*}

\begin{figure}

    \includegraphics[width=\columnwidth]{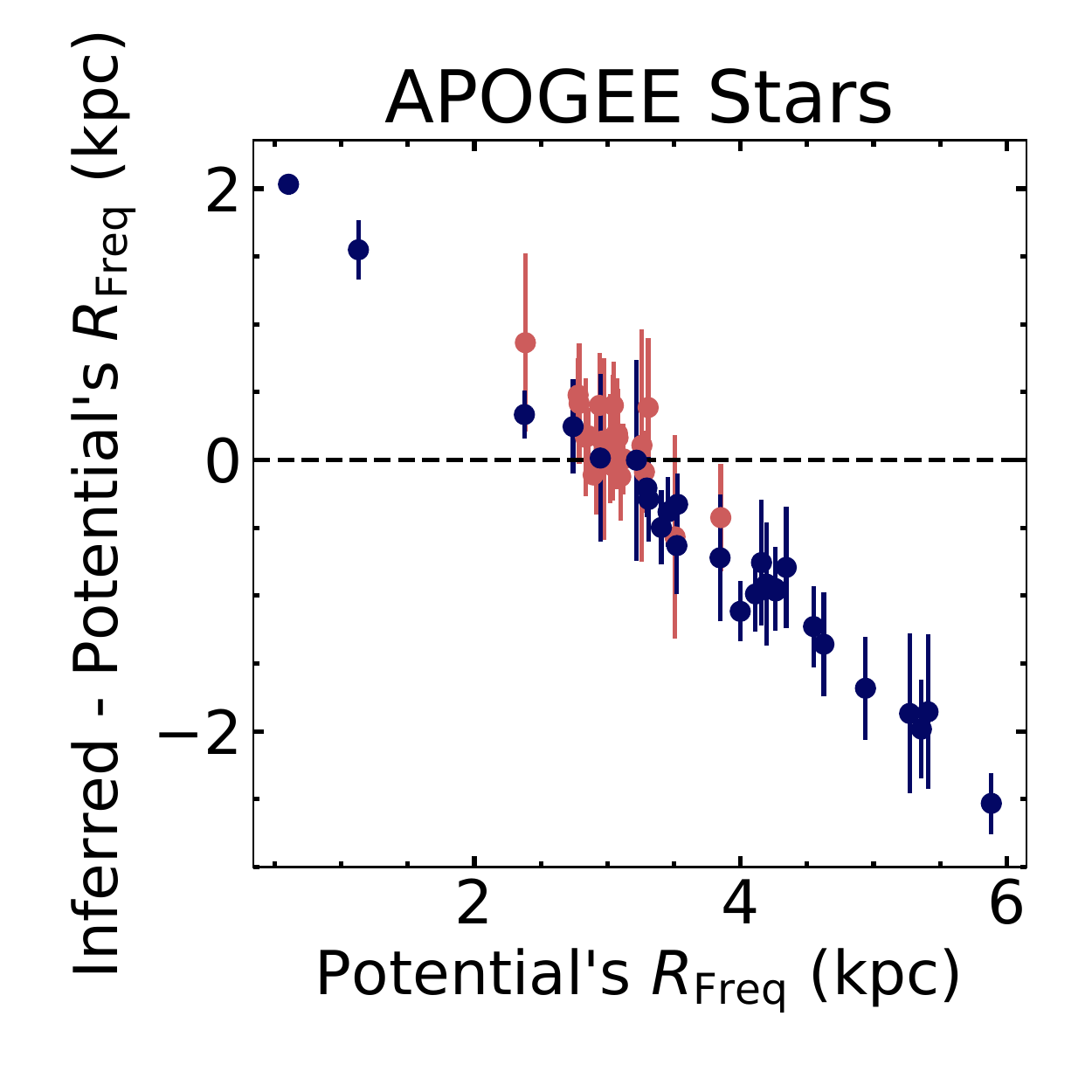}
    \caption{The comparison of the inferred \ro\ to the potential's \ro\ for APOGEE and \gaia\ stars that have been integrated in potentials with different bar lengths for 1 Gyr each, assuming a bar pattern speed of 41 km/s/kpc. Each point corresponds to the median inferred \ro\ of 10 random samples of $\approx$21,500 stars each with the error bar corresponding the standard deviation. The dark blue points correspond to potentials extracted from Galaxy A while the red points use potentials extracted from Galaxy B. We find potentials with \ro\ $\approx$3 kpc, which give the most self-consistent inferred \ro s (i.e., the difference between the inferred and potential's \ro\ is $\approx$0 kpc). One of these potentials is Model 1.
    }
    \label{fig:apo_len}
\end{figure}



Currently, the gravitational potential in the center of the Milky Way is poorly understood, partly because the length of the Galactic bar is not well-constrained. However, several studies assume potentials and make conclusions about the Galactic bar based on the resulting stellar orbits \citep[e.g.,][]{Queiroz2021,Lucey2021,Wylie2021}. Furthermore, it is difficult to constrain exactly how the deviations of the assumed potential from the true underlying mass distribution will impact the stellar orbits and therefore the conclusions drawn from them. In this work, we have already discovered that \ro\ inferred from the stellar orbits changes when the assumed gravitational potential and pattern speed of the bar changes. This demonstrates that any conclusions from stellar orbits are greatly impacted by the assumed gravitational potential, but we can use this result to our advantage by finding which potential gives a consistent result.

Using simulations, we have demonstrated that the inferred \ro\ from the apocenter distribution is only consistent with the \ro\ of the gravitational potential when that potential has a similar bar length to the snapshot from which the initial positions and velocities of the stars are extracted (see Figure \ref{fig:len}). Therefore, we can determine which potential is consistent with observed positions and velocities of stars by determining which potential gives a consistent inferred \ro\ measurement. We apply this to the Milky Way by integrating APOGEE and \textit{Gaia} stars in a variety of Milky Way-like potentials with different bar lengths. 

Following the same methods as in Section \ref{sec:method}, we integrate the orbits for 1 Gyr  in a variety of potentials and report the 99.5th percentile of the apocenter distribution of stars in the bar as our inferred \ro. At first, we assume a bar pattern speed of 41 km/s/kpc, consistent with previous estimates of the Milky Way's bar pattern speed \citep{Portail2017,Sanders2019,Bovy2019}. We calculate the orbits for the 215,869 stars shown in Figure \ref{fig:data}. We divide this sample into 10 random samples so that we are computing $\approx$21,500 orbits at a time, which is the same order of magnitude of the simulation samples we used in Section \ref{sec:method}. The final inferred \ro\ that we report is the median of the inferred \ro s from the 10 random samples with the standard deviation as the corresponding uncertainty.

In Figure \ref{fig:apoig}, we show the orbital frequency maps, galactic distribution and apocenter distributions of APOGEE/\textit{Gaia} stars integrated in the Model 1, 2, and 3 potentials. The left column of Figure \ref{fig:apoig} is similar to the left column of Figure \ref{fig:big} with the selection of bar stars to the left of the red-dashed line. The middle panel shows the Galactic distribution of all of APOGEE/\textit{Gaia} data with the selected bar stars shown in red. We show the same 5 kpc long bar as in Figure \ref{fig:data} rotated to 27$^{\circ}$ degrees in a dashed black line. In addition, we also show a circle with the radius equivalent to the potential's \ro\ derived from the apocenter distribution as a black solid line. We note that the Milky Way's bar angle is uncertain, thought estimates generally range from 25-27$^{\circ}$. From visual inspection of the center panels of Figure \ref{fig:apoig}, the stars selected as bar orbits appear to cluster closer to $Y\approx$0 kpc rather than higher $Y$ values, suggesting the bar angle may be smaller than 27$^{\circ}$. However, we defer determining whether the bar angle can be constrained from a similar method of checking for self-consistency until future work. 

In the right column of Figure \ref{fig:apoig}, we show the apocenter distribution of the disk stars in grey with the selected bar stars in red. The 99.5th percentile of the selected bar stars apocenter distribution (i.e., the inferred \ro), is shown as a red dashed line with the potential's \ro\ as a black solid line. The inferred \ro\ and potential's \ro\ are similar for the APOGEE/\textit{Gaia} data integrated in the Model 1 potential but are increasing different for the Model 2 and 3 potentials. This can also be seen from the Galactic distribution of the selected bar stars in the middle panel of Figure \ref{fig:apoig}. For the APOGEE/\textit{Gaia} data integrated in Model 1, the selected bar stars' Galactic distribution (red points) agrees with the potential's \ro\ (black dashed line). However, for the data integrated in the Model 2 and 3 potentials, the selected bar stars' distribution mostly ends well within the potential's \ro\, indicating that the potential's \ro\ is likely larger than the data's true bar length. Based on our previous results from simulations, Model 1 is more consistent with the APOGEE/\textit{Gaia} than Models 2 and 3 given that its potential leads to a more consistent \ro\ inference. However, it is important to test all of the 29 Galaxy A potentials and the 29 Galaxy B potentials to determine which of these potentials lead to the most consistent result.

In Figure \ref{fig:apo_len}, we show the difference between the inferred \ro\ and the potential's \ro\ for a variety of potentials with different bar lengths using APOGEE and \textit{Gaia} stars. The dark blue points correspond to potentials extracted from Galaxy A \citep{Bennett2021b} while the red points correspond potentials from Galaxy B \citep{Tepper-Garcia2021}. Each point is calculated using the median inferred \ro\ of 10 samples of $\approx$21,500 APOGEE and \textit{Gaia} stars with the uncertainty as the standard deviation. As demonstrated with simulations in Section \ref{sec:method}, the gravitational potentials that are most consistent with the APOGEE and \textit{Gaia} data will have a difference between the inferred \ro\ and the potential's \ro\ that is closest to zero kpc. From \ref{fig:apo_len}, we find that the potentials with a \ro\ of $\approx$3 kpc are most consistent with the APOGEE and \gaia\ data. Galaxy B has many potentials with \ro\ $\approx$3 kpc and are therefore consistent with the APOGEE and \gaia\ data. The Galaxy B potential that is most consistent with the APOGEE data has a potential \ro\ of 3.12 kpc. The inferred \ro\ from integrating the APOGEE data in this potential is 3.13 kpc, 0.01 kpc higher than the potential's. On the other hand, the most consistent Galaxy A potential is Model 1 with \ro\ of 3.22 kpc, and the resulting APOGEE inferred \ro is 3.22 kpc, only 0.002 kpc shorter than the potential's. 

We also recalculate Figure \ref{fig:apo_len} assuming a pattern speed of 60 km/s/kpc to match the estimate from \citet{Wang2012}. With this pattern speed, we find that all of the Galaxy A potentials give an inferred \ro\ that is much shorter than the potential's \ro. However, inferred \ro s measured using Galaxy B potentials are less impacted by the faster pattern speed. We find that the Galaxy B potentials with \ro$\approx$3 kpc are still the only potentials that give self-consistent inferred \ro s.

In this work, we have only tested potentials based on N-body snapshots. However, the method presented here can be used to test any Milky Way potential model, including analytical potentials. One would need to sample initial phase-space coordinates from the analytical potential distribution in order to calculate the potential's \ro.

It is important to note that our bar length estimate  does \textbf{not} reclassify the Galactic bar as a `fast' bar. The `fast'/`slow' classification is based on the dimensionless ratio $\mathcal{R} = R_{CR}/R_{bar}$ where $R_{CR}$ is the corotation radius and $R_{bar}$ is the bar length. Historically, these classifications are performed using estimates of the bar length that are based on the number density counts \citep{Debattista2000,Chiba2021} which is different than the dynamical \ro\ we measure above. For further discussion of  the various methods for estimating and defining  bar lengths see Section \ref{sec:discuss}.

\section{Methods of Measuring Bar Length}
\label{sec:discuss}

\begin{table}
\caption{Bar Length Estimates for Model 1 }
\label{tab:table1}
\begin{tabular}{cc}
\hline\hline
 & Model 1  \\
\hline
Method &  {Bar Length Estimate}  \\ 
& (kpc) \\
 \hline
\rf & 4.84  \\
Potential's $R_{\Omega}$ & 3.22 \\
Inferred $R_{\Omega}$ with APOGEE Stars  & 3.22 \\
\rx & 3.50 \\
\hline
\end{tabular}
\
\end{table}

In this work, we define a quantity, \ro, as a measure of bar length (see method described in Section \ref{sec:method}). However, \ro\ is not a robust measurement of the bar length since it excludes $x_1$ orbits which have $\Omega_y/\Omega_x$=1 and are the backbone of most bars \citep{Wang2016}. There are a number of methods to measure bar lengths, but each method can lead to varying results. The dynamical length of a bar is defined by the maximal extent of trapped orbits. As $x_1$ orbits are the dominant bar orbits, especially at the ends, their maximal extent is the dynamical length of a bar. In this section, we determine the maximal extent of $x_1$ orbits for the Model 1 potential and compare it to other methods of measuring the length of bars. We focus on the Model 1 potential for this comparison, as it was the most consistent with the APOGEE/\gaia\ data. However, we note that the consistency of bar length measurements using different methods can be very dependent on the bar age and morphology \citep{Petersen2019b,Hilmi2020}.

One common method for measuring the bar length in external and model galaxies is Fourier decomposition using $A_{2,2}/A_0$ (see Section \ref{sec:data} for the calculation). For our Model 1 potential this leads to a bar length measurement of 4.84 kpc.  However, the Fourier decomposition method has been shown to measure larger bar lengths than other methods \citep{Petersen2019b}. Similar to the method for measuring the bar length in external galaxies by fitting ellipses to the surface brightness, the Fourier decomposition method especially measures longer bar lengths when spiral arms are attached which is the case for most of the bars in external galaxies \citep{Hilmi2020}. Model 1  has weak spiral arms attached to the bar so it is possible that the measurement of bar length with this method is especially large. Interestingly, the \rf\ of 4.84 kpc for this potential is similar to the Milky Way's bar length estimate (5 kpc) from number density counts \citep{Wegg2015}. However, \citet{Hilmi2020} suggest this measurement may be overestimated by 1-1.5 kpc given recent observations of spiral arms attached to the bar \citep{Rezaei2018}. 

We also use another dynamical method of measuring the bar length from \citet{Petersen2021}. For a complete description of the method we refer the reader to \citet{Petersen2016,Petersen2021}. In short, the method classifies bar stars based primarily on the angular distance between the apocenter positions of the star's orbit and the bar axis.  Using this method we can cleanly select $x_1$ orbits which is the family of orbits associated with the inner Lindblad resonance ($2\Omega_{\phi}-\Omega_r=1\Omega_{\rm{b}}$) and whose maximal extent provides a robust estimate of the dynamical length of the bar \citep{Petersen2019b}.  In similar models, \citet{Petersen2019a} found that the $x_1$ orbits are responsible for nearly all of the self-gravity of the $m=2$ Fourier mode bar. With the $x_1$ orbits we measure \rx= 3.50 kpc for Model 1, which is slightly longer than the potential's \ro, consistent with expectations that $x_1$ orbits make up the longest part of the bar \citep{Wang2016}.  

Given the variation of measured bar lengths from the different methods, it is important to be careful when comparing reported bar length estimates in the literature. To avoid this, instead of emphasizing  a specific bar length measurement, we emphasize Model 1 as the most consistent model for the Milky Way bar, for which we measure a \textit{dynamical}\footnote{We define the dynamical bar length as \rx, the maximal extent of trapped $x_1$ orbits that participate in the solid-body rotation.} bar length of \rx $\approx$3.5 kpc with an overdensity that extends to \rf $\approx$4.8 kpc. We encourage a movement towards publicly available potential models that would allow for easier direct comparison between dynamical results for the inner Galaxy. We note our method can be used to check any potential for consistency with Milky Way data. However, it is important to test any potential model with simulated particles, as we have done in this work, to ensure a bias is not introduced by an incomplete selection of bar orbits with the orbital frequency method.

Another important thing to note is the dependence of the orbital structure of the inner Galaxy on the assumed potential model. This is apparent from looking at the orbital frequency maps in Figure \ref{fig:apoig}. Using the three different potential models, we find the distribution of fundamental frequencies for the orbits are significantly different. The distribution of orbital frequencies for Model 1 and 2 are somewhat similar, but the ratios of the longest distance from the Galactic center along the bar's major axis to the bar's minor axis ($x_{max}/y_{max}$) are quite different. Although we have found that Model 1 leads to the most consistent result for the \ro\ determination, we are unsure if other parameters (e.g, the vertical structure and width) are accurate approximations for the Milky Way. Furthermore, we are unsure of how these other parameters may impact the orbital structure. Therefore, we caution the community to be wary when making conclusion about the inner Galaxy from orbits without doing a thorough investigation on the potential dependence.

\section{Conclusions}
\label{sec:conclu}

In this work, we develop a new method to dynamically estimate the bar length directly from orbit integration. First, we verify this method using simulations. In short, we select a sample of bar stars using fundamental frequency maps of orbits and use the 99.5th percentile of the apocenter distribution as the \ro. We find that when the initial position and velocity distribution of the star particles are extracted from the distribution of the potential model, then we achieve a consistent result in that the measured \ro\ from the orbits matches that of the potential. However, if the initial positions and velocities are extracted from a significantly different distribution than the potential's, then the measured \ro\ from the orbits is different than that of the potential used to calculate said orbits. With this result, we can find the potential  that matches the distribution from which the initial positions and velocities are taken by finding which potential leads to a self-consistent measured \ro. 

We then apply this new method to the Milky Way and find which potential leads to a sellf-consistent measured \ro\ for the APOGEE/\gaia\ data. We find our Model 1  (see Section \ref{sec:method} and Figure \ref{fig:a2}) is the most consistent potential.  This model is derived from the \texttt{MWPotential2014-3} simulation in \citet{Bennett2021b} and has a dynamical bar length of 3.50 kpc. 

However, it is important to note that there are many methods of estimating the bar length, which can lead to biases dependent on bar age and morphology \citep{Petersen2019b,Hilmi2020}. For the Model 1  potential, we also measure a bar length of 4.84 kpc from the $m=2$ mode of the Fourier decomposition. However, this method is known to give high estimates of bar lengths, especially in the case of connecting spiral arms. Given the inconsistency of bar length measures, we emphasize the importance of making gravitational potential models public for the Milky Way in order to make fair comparisons between results. Furthermore, we note that our method for checking consistency to constrain the bar length can be used with any potential and we encourage the community to test their favorite potential before using it to draw conclusions about the inner Milky Way. In future work, we plan to further investigate the orbital structure of the Milky Way's bar and its dependence on various potential model parameters.

\section*{Acknowledgements}
{\small 
This work was developed at the Big Apple Dynamics School and the Pre-Doctoral Program in 2021 at the Flatiron Institute. We thank them for their generous support. 

This material is based upon work supported by the National Science Foundation Graduate Research Fellowship under Grant No. 000392968. Support for SP was provided by NASA through the NASA Hubble Fellowship grant \#HST-HF2-51466.001-A awarded by the Space Telescope Science Institute, which is operated by the Association of Universities for Research in Astronomy, Incorporated, under NASA contract NAS5-26555. KH acknowledges support from the National Science Foundation grant AST-1907417 and AST-2108736 and from the Wootton Center for Astrophysical Plasma Properties funded under the United States Department of Energy collaborative agreement DE-NA0003843. This work was performed in part at Aspen Center for Physics, which is supported by National Science Foundation grant PHY-1607611. This work was performed in part at the Simons Foundation Flatiron Institute’s Center for Computational Astrophysics during KH’s tenure as an IDEA Fellow. MSP is partially supported by grant Segal ANR-19-CE31-0017 of the French Agence Nationale de la Recherche (https://secular-evolution.org).

Funding for the Sloan Digital Sky Survey IV has been provided by the Alfred P. Sloan Foundation, the U.S. Department of Energy Office of Science, and the Participating Institutions. SDSS acknowledges support and resources from the Center for High-Performance Computing at the University of Utah. The SDSS web site is www.sdss.org.

SDSS is managed by the Astrophysical Research Consortium for the Participating Institutions of the SDSS Collaboration including the Brazilian Participation Group, the Carnegie Institution for Science, Carnegie Mellon University, Center for Astrophysics | Harvard \& Smithsonian (CfA), the Chilean Participation Group, the French Participation Group, Instituto de Astrofísica de Canarias, The Johns Hopkins University, Kavli Institute for the Physics and Mathematics of the Universe (IPMU) / University of Tokyo, the Korean Participation Group, Lawrence Berkeley National Laboratory, Leibniz Institut für Astrophysik Potsdam (AIP), Max-Planck-Institut für Astronomie (MPIA Heidelberg), Max-Planck-Institut für Astrophysik (MPA Garching), Max-Planck-Institut für Extraterrestrische Physik (MPE), National Astronomical Observatories of China, New Mexico State University, New York University, University of Notre Dame, Observatório Nacional / MCTI, The Ohio State University, Pennsylvania State University, Shanghai Astronomical Observatory, United Kingdom Participation Group, Universidad Nacional Autónoma de México, University of Arizona, University of Colorado Boulder, University of Oxford, University of Portsmouth, University of Utah, University of Virginia, University of Washington, University of Wisconsin, Vanderbilt University, and Yale University.

This research made use of Astropy,\footnote{http://www.astropy.org} a community-developed core Python package for Astronomy \citep{astropy:2013, astropy:2018}. Other software used includes \textit{IPython} \citep{ipython}, \textit{matplotlib} \citep{matplotlib}, \textit{numpy} \citep{numpy}, \textit{galpy}\footnote{http://github.com/jobovy/galpy} \citep{Bovy2015}, \textit{gala} \citep{gala,adrian_price_whelan_2020_4159870}, \textit{AGAMA} \citep{Vasiliev2019}, and \textit{scipy} \citep{scipy}.

This work has made use of data from the European Space Agency (ESA) mission {\it Gaia} (https://www.cosmos.esa.int/
gaia), processed by the {\it Gaia} Data Processing and Analysis Consortium (DPAC, https://www.cosmos.esa.int/web/gaia/dpac/
consortium). Funding for the DPAC has been provided by national institutions, in particular the institutions participating in
the Gaia Multilateral Agreement.

}

\section*{Data Availability}
{\small
The data underlying this article were accessed from an SDSS Value Added Catalog which can be found at \url{https://data.sdss.org/sas/dr17/env/APOGEE_ASTRO_NN/}. The derived data generated in this research will be shared on reasonable request to the corresponding author. The Galaxy A simulation will be made publically available upon acceptance. The Galaxy B simulation is already available at \url{http://www.physics.usyd.edu.au/mwbar/sim_data/}.
}
\bibliography{bibliography}
\bsp	
\label{lastpage}
\end{document}